# Effect of Mn substitution on the delicate balance between structure and properties of $Gd_{1.4}Ce_{0.6}Sr_2RuCu_2O_{10}$

S. Kalavathi, J. Janaki, T. N. Sairam, Awadhesh Mani, R. Rawat[+] and V. Sankara Sastry

Materials Science Division, Indira Gandhi Centre for Atomic Research, Kalpakkam 603102

+ UGC-DAE Consortium for Scientific Research, University Campus, Khandwa Road. Indore-452017

**Abstract**

Some new members of a Ruthenocuprate (2212) series have been synthesized by Mn substitution for Ru in $Gd_{1.4}Ce_{0.6}Sr_2RuCu_2O_{10}$. Characterisation by XRD phase analysis followed by Rietveld refinement has been carried out. IR spectroscopy studies and Rietveld refinements are consistent with substitution of Mn at Ru site. XRD studies indicate changes in structural features on substitution including a significant change in lattice parameter for a very low substitution level (1mole% of Ru). The pristine compound shows coexistence of superconductivity and magnetism. Four probe resistivity studies indicate a semiconductor like upturn in resistivity and absence of superconductivity even for Mn substitution levels as low as 1 mole%. a. c. susceptibility measurements show a progressive suppression of the magnetic transition temperature as well as a smearing of the magnetic transition as a function of Mn substitution. Possible reasons for absence of superconductivity have been discussed.

**INTRODUCTION**

Coexistence of superconductivity and ferromagnetism has been a subject matter of considerable interest as these two phenomena are believed to be mutually inimical. Ruthenocuprates hold a special position amongst the materials exhibiting such a coexistence since in these materials magnetic transition precedes the superconducting transition. For instance magnetic transition occurs at 130K ($T_{M1}$) and 95K ($T_{M2}$) in the layered compound $Gd_{1.4}Ce_{0.6}Sr_2RuCu_2O_{10}$ (1222) and the superconducting transition ($T_c$) occurs at 40 K. Observation of hysteresis in magnetization measurements at temperatures around 5 K is considered as a confirmation of the existence of ferromagnetism [1] in the superconducting state. A number of studies have been made on Ruthenocuprates to identify the origin of the two magnetic transitions and the co-existing superconducting transition. A homogenous superconducting state with spontaneous vortex state has been put forward as one possibility [2,3]. Phase separation in the normal state in terms of antiferromagnetic regions sprinkled with weakly ferromagnetic regions has been put forth as another possibility[4]. According to the latter superconductivity itself results at low temperature from the antiferromagnetic regions.

Ruthenocuprates of type (1222) e.g., $Gd_{1.4}Ce_{0.6}Sr_2RuCu_2O_{10}$ exhibit a tetragonal structure with space group $I4/mmm$ [5,6]. The structure can be derived from $YBa_2Cu_3O_{7-\delta}$ structure by the insertion of a fluorite type $(Ln,Ce)O_2$ layer between the bases of the $CuO_5$ pyramids. The latter shifts the alternate perovskite blocks by $a+b/2$. It is widely accepted that the $CuO_2$ layer is responsible for superconductivity and the $RuO_2$ layer contributes to magnetism. Unlike the $YBa_2Cu_3O_{7-\delta}$ where the Cu-O chains act as charge reservoirs, in this compound there are two possible charge reservoir layers namely the $(Ln,Ce)O_2$ and $RuO_2$ layers. Our earlier studies indicate that Dy substitution at $(Ln,Ce)O_2$ site affects both the superconducting and magnetic transition temperatures[6]. It is known from literature that among $Gd_{2-x}Ce_xSr_2RuCu_2O_{10}$ the compound with composition $Gd_{1.4}Ce_{0.6}Sr_2RuCu_2O_{10}$ shows the highest of $T_c$ [7]. It is also known that the $T_c$ and the broadening of transition depends on the synthesis route namely, the compounds synthesized at high pressure oxygen environment show better properties[8] in terms of $T_c$, $H_{c2}$ and so on in comparison with the ambient pressure oxygen annealed samples. Annealing in air retains magnetic transitions but destroys the superconductivity implying the role of oxygen in hole doping the $CuO_2$ layers. The valency of Ru has been found to lie between 4.95 and 5 as seen from XANES measurement [9]. Cationic substitution of the Ru site either completely by Co, Fe and Mo[10,11] or partially by Zn[12] have been studied in literature . In most cases, both $T_c$ and $T_M$ deteriorate on substitution. The only system that retains superconductivity after substitution at the Ru site is $MoSr_2R_{1.5}Ce_{0.5}Cu_2O_{10-\delta}$ [11]. In this case for R= Sm and Eu the $T_c$ is 13 and 26 K respectively and for heavy R elements Ho- Lu the pentavalent Mo layers are antiferromagnetically ordered with $T_M$ ranging from 13-27 K. Hence it is inferred that there is a competition between superconductivity and magnetism rather than co-existence in these systems [11]. In the present work substitution at Ru site by manganese has been carried out to study the role of Mn in modifying the magnetic and superconducting transitions of the $Gd_{1.4}Ce_{0.6}Sr_2RuCu_2O_{10}$ A series of samples have been synthesized with Mn concentrations of 1, 2, 5, 10 and 30 mole % substituted for Ru in $Gd_{1.4}Ce_{0.6}Sr_2RuCu_2O_{10}$. Studies on these samples have shown interesting observations.

## EXPERIMENTAL

Predominantly single phase polycrystalline samples of $Gd_{1.4}Ce_{0.6}Sr_2Ru_xMn_{1-x}Cu_2O_{10}$ (<5% impurity phases) have been synthesized starting from high purity ( > 99.9% ) Gd2O3, CeO2, SrCO3, CuO and MnO2 by solid state reaction at 1000 °C for 24 hrs. followed by repeated regrinding and oxygen annealing at 1050° C for 48 hrs. XRD characterization has been carried out on a STOE diffractometer with Si (911) zero back-ground plate in the Bragg Brentano geometry. Slow scan XRD data have been acquired on pristine and x= 0.3 samples with 0.02 step size and an acquisition time of 36 hrs. Rietveld refinement using GSAS-EXPGUI software[13] has been carried out for x=0 and x=0.3 XRD data. Peak shapes have been treated assuming a Pseudo Voigt profile . The refinement has been carried out in the following sequence: scale factor, background parameter, zero shift, cell parameters, profile parameters, positional parameters ,site occupancies and thermal parameters. IR measurements in the near-normal incidence reflection geometry were performed on sintered pellets using BOMEM DA8 FTIR spectrometer operating at a resolution of 4 cm$^{-1}$. A gold mirror was used as the reference. Kramers-Kronig analysis on reflectance data was done to get the complex dielectric constant from which the real part of the conductivity was obtained using the relation $\sigma(\omega)=(\omega/4\pi) \varepsilon_2(\omega)$. XPS measurements were carried out on powder samples of Mn composition x=0 and x=0.3 using VG ESCA LAB MK 200 system with Al K-alpha x-ray source and hemispherical analyser. Room temperature thermopower measurements have been carried out in a home made thermopower apparatus. Four probe resistivity measurements have been performed from 4.2 to 300 K. Relative magnetic susceptibility measurements have been done using an home made dipstick type apparatus operating at 941 Hz and 0.25 Oe .

## RESULTS

XRD phase analysis indicates that the system remains as single phase up to x= 0.3. The d- spacing, intensities and the Miller indices for the pristine and the end member of the series namely $Gd_{1.4}Ce_{0.6}Sr_2Ru_{0.7}Mn_{0.3}Cu_2O_{10}$ have been reported in the ICDD Powder data file 2005[14]. For x>0.3, XRD indicates multiphasic nature of the system and complete substitution of Ru by Mn is not possible as it leads to the formation of $Sr_3Mn_2O_7$ as a major phase and impurity phases $SrCuO_2$, $CeO_2$ and CuO. Mn substitution in the lattice and formation of the new member could be well characterized by the change in lattice parameters for Mn up to x=0.3. The tetragonal lattice parameters shift systematically from a=3.835Å and c=28.570 Å for the pristine compound to a= 3.839 Å and c=28.511 Å for the Mn substituted compound with x=0.3. The most striking observation with respect to the structural features is the significant change in c lattice parameter for a concentration of Mn as low as 1% (Fig 1a). The decrease in c- parameter on Mn substitution is more than what would be expected due to the smaller ionic radius alone ( Ionic radius of $Mn^{+4}$, VI coordination = 0.54 Å as compared to 0.565 Å for $Ru^{+5}$, VI coordination and 0.63 Å for $Ru^{+4}$, VI coordination ). It may therefore imply a change in the valence state of Cu and a consequent relaxation in the atomic positions.

Rietveld refinement indicates residual values (Rp) less than10% in agreement with X-ray Rietveld refinement analysis reported in literature using sealed tube X-ray source [15]. Fig 2 shows the results of Rietveld refinement of the powder XRD pattern of $Gd_{1.4}Ce_{0.6}Sr_2Ru_{1-x}Mn_xCu_2O_{10}$ (x=0.3) .O1, O2, O3 and O4 represent the four oxygen sites namely the apical oxygen, $RuO_2$ planar oxygen, $CuO_2$ planar oxygen and oxygen corresponding to the $Gd(Ce)O_2$ block respectively. Comparison of refined data between x=0 and x=0.3 (Table 1) shows a reduction in the apical Ru-O1 bond length (around 3%) and an increase in the apical Cu-O1 bond length. This is consistent with Mn substitution at the Ru site and possible change in charge transfer between the $RuO_2$ and $CuO_2$ planes. Attempts were made to refine Mn occupancy by considering two models one in which it occupies the Cu site and the other in the Ru site. The latter gave a refinement residual value slightly lower than the former implying Mn occupying the Ru site

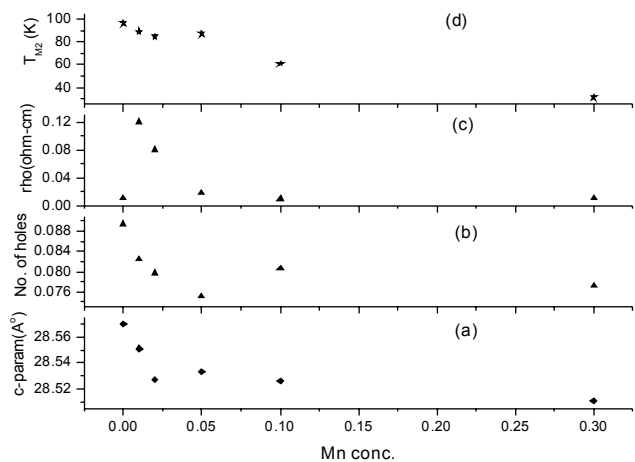

Fig.1 Variation of (a) c-parameter, (b) No.of holes per Cu atom (c) room temperature resistivity and (d) magnetic transition temperature ($T_{M2}$) as a function of Mn conc. in $Gd_{1.4}Ce_{0.6}Sr_2Ru_{1-x}Mn_xCu_2O_{10}$ (x=0, 0.01, 0.02, 0.05, 0.1, 0.3)

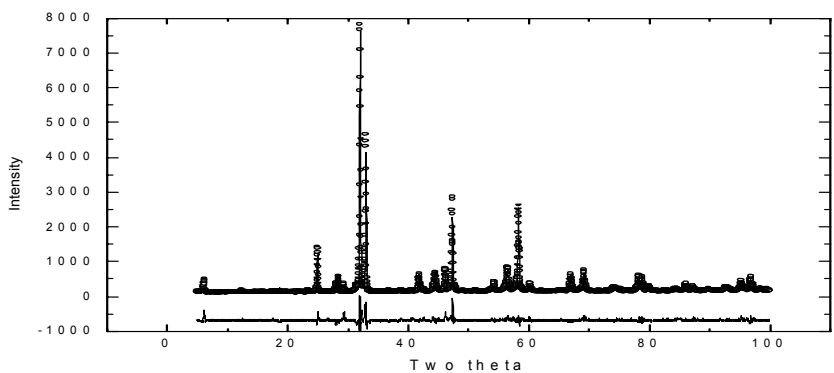

Fig.2 Rietveld fit to the XRD data of $Gd_{1.4}Ce_{0.6}Sr_2Ru_{0.7}Mn_{0.3}Cu_2O_{10}$

Table 1 : Refined positional parameters and bond distances of $Gd_{1.4}Ce_{0.6}Sr_2Ru_{1-x}Mn_xCu_2O_{10}$ (x=0,0.3)

| Atom | Site | Mn=0 | | | Mn=0.3 | | | Bond | Mn=0 | Mn=0.3 |
|---|---|---|---|---|---|---|---|---|---|---|
| | | x | y | z | x | y | z | Ru-O1 | **1.974** | **1.939** |
| Gd/Ce | 4e | 0.5 | 0.5 | 0.205 | 0.5 | 0.5 | 0.205 | Ru-O3 | 1.978 | 2.131 |
| Sr | 4e | 0.5 | 0.5 | 0.078 | 0.5 | 0.5 | 0.078 | Mn-O1 | 1.974 | 1.939 |
| Ru/Mn | 2a | 0 | 0 | 0 | 0 | 0 | 0 | Mn-O3 | 1.978 | 2.131 |
| Cu | 4e | 0 | 0 | 0.144 | 0 | 0 | 0.142 | Cu-O1 | **2.145** | **2.174** |
| O1 | 16n | 0.021 | 0 | 0.069 | 0.09 | 0 | 0.067 | Cu-O2 | 1.926 | 1.926 |
| O2 | 8g | 0 | 0.5 | 0.150 | 0 | 0.5 | 0.137 | | | |
| O3 | 8j | 0.125 | 0.5 | 0 | 0.24 | 0.5 | 0 | Cu-O2-Cu | 170.1 | 171.5 |
| O4 | 4d | 0 | 0.5 | 0.5 | 0 | 0.5 | 0.5 | | | |

Fig. 3 shows the optical conductivity spectra of Ru 2212 carried out [16] for various Mn substitution. The phononic part of the of the optical spectra, obtained after Kramers-Kronig transformation, shows prominent optical modes at 135, 152, 184, 291, 522 and 672 cm$^{-1}$ corresponding to vibrations of Cu, Gd, Ru, O2, O3 and apical oxygen atom (O1) respectively. The mode identification has been done by comparing with the IR spectrum of $GdSr_2RuCu_2O_8$ [17]. On substitution of Mn, both the Cu and Gd modes seem to be unaffected, but the Ru mode frequency, $\omega$, shows a noticeable change. In addition, the mode corresponding to the apical oxygen atom shows a red shift on substituting Gd2122 with Mn.

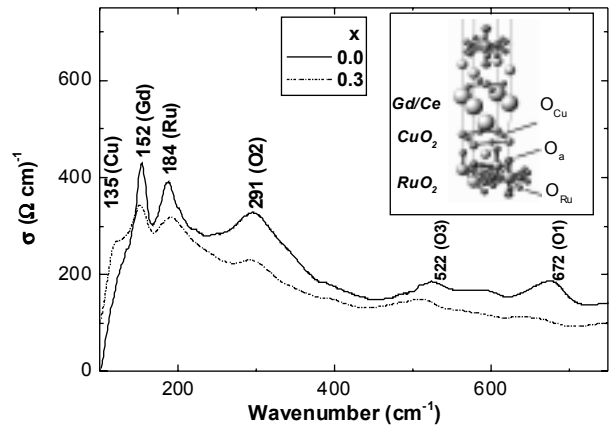

Fig.3 IR spectra of $Gd_{1.4}Ce_{0.6}Sr_2Ru_{1-x}Mn_xCu_2O_{10}$ (x=0, 0.3) showing optical conductivity as a function of wave number

It is believed that in the rutheno cuprates the Cu-$O_2$ layers are responsible for superconductivity and Ru-$O_2$ layers for magnetism and hence study of the valence state of the cations would be important. It is already known from literature that Ru valency itself varies between 4.95 and 5 [18] in Ru(1222) system. Therefore Cu XPS measurements have been carried out on pristine and x=0.3 Mn substituted sample to know about the valence state of Cu.

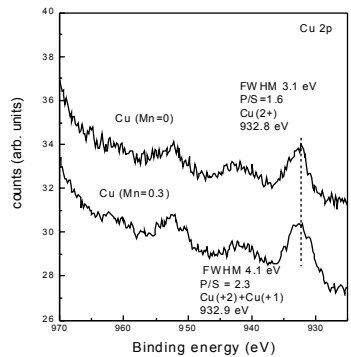

Fig .4 Cu 2p core level XPS spectra of $Gd_{1.4}Ce_{0.6}Sr_2Ru_{1-x}Mn_xCu_2O_{10}$ (x=0, 0.3)

From the results shown in Fig (4) it is seen that two main peaks characteristic of Cu 2p states are observed as expected for $Cu^{2+}$[19]. The first $Cu^{2+}$ peak appears at 932.8 eV along with the satellite peak for the pristine sample. In the case of Mn substituted sample with x=0.3 the peak width increases to 4.1 eV and the peak to satellite ratio also has increased from 1.6 to 2.3. This is characteristic of [19] reduction of Cu atoms from $Cu^{2+}$ to $Cu^{1+}$. We believe that this may be related to the decrease in charge carrier concentration in Cu-O plane upon Mn substitution in the Ru plane.

Results of four probe resistivity are presented in Fig (5). In the case of the pristine sample the four probe resistivity measurement shows onset of superconducting transition at 32 K. This $T_c$ value is consistent with $T_c$ value for samples not subjected to high pressure oxygen annealing. With just 1% substitution by Mn, superconductivity is completely destroyed in the sample and a steep increase in resistivity results down to 4 K. For Mn substitution beyond 1%, all samples show a semiconductor like increase in resistivity. Room temperature resistivity (Fig (1c)) is found to increase by an order of magnitude for the 0.01 Mn substituted sample. It decreases thereafter and saturates for higher Mn substitutions up to 0.3. Room temperature thermopower of the pristine sample is found to be 32.75 μV/K. It rises steeply to 42.75 μV/K for the 0.01 Mn substituted sample and thereafter increases slowly. The hole concentration can be derived from thermopower using the relation S(290K)= 992*exp (-38.1* p) [20]. The variation of hole concentration as a function of Mn is shown in fig. (1b).

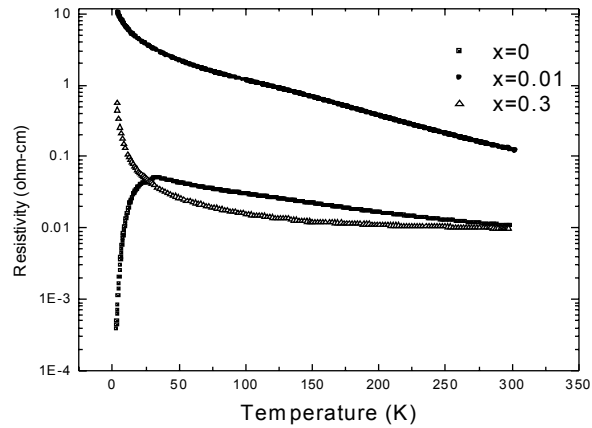

Fig.5 Four probe Resistivity vs Temperature for $Gd_{1.4}Ce_{0.6}Sr_2Ru_{1-x}Mn_xCu_2O_{10}$ (x=0, 0.01, 0.3)

The a.c susceptibility measurement Fig. 6 confirms superconductivity of the pristine sample at 29 K. Two magnetic transitions are observed at 130 K ($T_{M1}$) and 96 K ($T_{M2}$). Inset to Fig. 6 shows presence of $T_{M1}$ in pristine sample and the absence of this magnetic transition on 1% Mn substitution. This transition is also absent for all other Mn concentrations. It is also observed that $T_{M2}$ is shifted to lower temperatures (Fig.1d) with increase in Mn concentration. Apart from this shift to lower temperatures, a smearing of the transition is also observed as a function of Mn concentration.

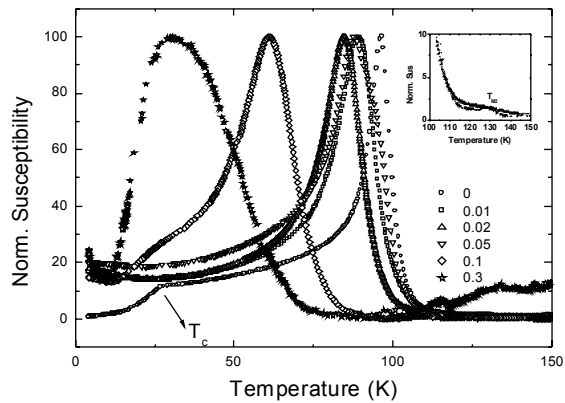

Fig.6 a.c. susceptibility vs Temperature for $Gd_{1.4}Ce_{0.6}Sr_2Ru_{1-x}Mn_xCu_2O_{10}$ (x=0, 0.01, 0.02, 0.05, 0.1, 0.3)

## DISCUSSION

It is known that the magnetic and superconducting properties observed in Ru(1222) system are a result of interactions in the Ru-O and Cu-O planes respectively. Also the interaction between the planes have a role to play in determining the properties of the system. There is evidence in literature [5] for the role of oxygen stoichiometry in affecting the structure and properties. While substitution of ions at Ru and Gd sites have been carried out in the past [10,11,12], our present work is the first to observe an anomalous change in physical properties and structural parameters for very small doping levels (1%). While no detailed study exists of the magnetic transition $T_{M1}$ observed at about 130 K, studies exist on $T_{M2}$. Recent neutron diffraction studies at low temperature (5K) of the compound $RuSr_2Nd_{0.9}Y_{0.2}Ce_{0.9}Cu_2O_{10}$ [21] indicate AFM ordering of Ru and Cu moments and a 45° canting of both these moments towards a parallel spin structure. X-ray absorption and NMR studies indicate mixed valence of Ru namely 95% $Ru^{5+}$ and 5% $Ru^{4+}$. Due to this mixed valence a double exchange mechanism inside the $RuO_2$ sheets was proposed, mediated by the small canting of the Ru octahedra as an explanation for the weak ferromagnetism[18]. The mixed valence also leads one to believe that the $RuO_2$ planes should be conducting as has been established experimentally from transport and spectroscopic studies [18]. Since the Mn doping has been carried out in oxygen atmosphere and a drastic decrease in c-parameter is observed it would be appropriate to believe that Mn enters either as $4^+$ or $5^+$ or as a mixture of $4^+$ and $5^+$ in the system. Detailed XAFS or XANES measurement can lead to a better understanding of the oxidation state of Mn. It is also clear that Mn will not assume $2^+$ and $3^+$ as they will have higher ionic radii and would not be consistent with the observed large decrease in c-parameter. If Mn enters as a mixture of $4^+$ and $5^+$ and occupies Ru, it would lead to a change in local co-ordination from octahedral to tetrahedral as $Mn^{5+}$ generally adopts a tetrahedral coordination [22]. This would also lead to a depletion of oxygen in the system thereby reducing a part of $Cu^{2+}$ to $Cu^{1+}$. This is in accordance with our XPS observation. This reduction in charge carrier concentration in $Cu-O_2$ plane destroys superconductivity even at 1% substitution and also drives the system to an insulating state. The relatively large fall in hole concentration with smaller percentage of Mn substitution which stabilizes later with higher Mn concentration is also observed in our thermopower data shown in Fig (1b). The decrease in resistivity on further substitution of Mn can be understood as follows: as Mn $5^+$ corresponds to $d_2$ configuration amidst $Ru5^+$ in $d_3$ configuration, this can probably bring about a hopping conductivity. Finally the lowering of the magnetic transition temperature may be because of the disorder caused in the Ru spin arrangement due to the presence of $Mn5^+$ with d2 configuration at random Ru sites. This can explain the smearing of $T_{M1}$ and it may also distort the canting of the Ru moments at lower temperatures thereby reducing the second magnetic transition temperature ($T_{M2}$).

## CONCLUSION

From the above results it is clear that Mn substitution in the system is possible up to x=0.3 and this leads to a significant change in lattice parameters. Mn occupies Ru location. The fact that Mn substitution in Ru has affected superconductivity implies that the Cu-O and Ru-O planes are coupled. The results of resistivity & ac susceptibility measurements indicate Mn substitution of even 1% destroys superconductivity and results in a semiconducting behaviour. It is conjectured that this could probably be due to Mn entering the Ru-O layer partially as 5+ causing a locally different co-ordination and leading to reduction of the charge carrier concentration in the $CuO_2$ planes. The major imbalance in property and structure on Mn substitution relative to the pristine sample corroborates this view. The disorder in the $RuO_2$ planes might modify the spin alignment and weaken the magnetic interaction leading to a reduction in the magnetic ordering temperature and a smearing of the peak.

## ACKNOWLEDGEMENTS

The authors wish to acknowledge Dr. Santanu Bera of Water and Steam Chemistry Division, IGCAR for providing XPS, Dr.Gayathri Banerjee, Materials Science Division, IGCAR for providing room temperature thermopower data and Dr. Manas Sardar Materials Science Division, IGCAR for fruitful discussions.